\documentclass[prl,twocolumn,aps,showpacs]{revtex4} % showkeys

\usepackage{graphicx}% Include figure files
\usepackage{dcolumn}% Align table columns on decimal point
\usepackage{bm}% bold math

\begin{document}

\title{Pruning the Tree of Life:\\
       $k$-core Percolation as Selection Mechanism}

\author{Peter Klimek$^{1}$, Stefan Thurner$^{1,2,*}$, Rudolf Hanel$^1$}
%\email{thurner@univie.ac.at} 

\affiliation{
$^1$ Complex Systems Research Group; HNO; Medical University of Vienna; 
W\"ahringer G\"urtel 18-20; A-1090; Austria \\
$^2$ Santa Fe Institute; 1399 Hyde Park Road; Santa Fe; NM 87501; USA \\
$^*$ E-Mail: thurner@univie.ac.at
} 

\begin{abstract}
%300 words
We propose a model for evolution aiming to reproduce statistical features of fossil data, in particular the distributions of extinction events, the distribution of species per genus and the distribution of lifetimes, all of which are known to be of power law type. The model incorporates both species-species interactions and ancestral relationships. The main novelty of this work is to show the feasibility of $k$-core percolation as a selection mechanism. We give theoretical predictions for the observable distributions, confirm their validity by computer simulation and report good agreement with the fossil data. A key feature of the proposed model is a co-evolving fitness landscape determined by the topology of the underlying species interactions, ecological niches emerge naturally. The predicted distributions are independent of the rate of speciation, i.e. whether one adopts an gradualist or punctuated view of evolution.

\end{abstract}

\maketitle

\section{Introduction}

A promising line of interdisciplinary research was initiated when it became apparent 
that a series of  statistical facts in various types of fossil data can not be obtained by 
a straight forward extension of existing ways of describing species proliferation and extinction 
over geological timescales.
The recent quantitative interest in evolutionary models 
%predicting the critical behaviour 
originates  from the fact that fossil data from different sources \cite{Willis22, Sepkoski92} shows 
power law behaviour with typical exponents for three observables: 
(i) the distribution of sizes of extinction events, (ii) the lifetime of species and (iii) the number of species per genus, 
see e.g. \cite{Newman03} for an overview. 

%Here we propose a model of evolution which synthesises and advances previous models and is able 
%to explain trends in the fossil data by the combination of two simple mechanisms.

One of the first quantitative models of evolution was the $NK$ model proposed by Kauffman 
\cite{Kauffman93}, where species evolve and compete on a rugged fitness landscape. A species' 
fitness and therefore lifetime is given by its genome and the randomly associated fitnesses to 
the respective genes. In a similar vein Bak and Sneppen \cite{Bak93} refined Kauffman's ideas to a model 
exhibiting self-organized criticality. Here it is assumed that the fitness landscape 
possesses valleys and peaks and over time a species will mutate "across" a fitness barrier to an 
adjacent peak. In contrast to these models, where there is no explicit species-species interaction, 
Sol\'e and Manrubia \cite{Sole96} constructed a model focusing on interspecies dependencies. 
They incorporate a connection matrix containing the mutual support between two species. 
If this support drops below a critical value the species will not be able to maintain its existence 
anymore, it will go extinct. In contrast to the $NK$ and Bak-Sneppen model which are 
\textit{per se} critical, the Sol\'e-Manrubia model's criticality is parameter dependent. 
For a review %of these models and some refinements and extensions of them 
see again \cite{Newman03}.

Recently a more general and abstract framework to treat systems subject to evolution was 
developed out of the notion of catalytic sets on networks 
%combined with combinatorics 
\cite{Hanel05, Hanel07}. 
We will use this approach to model species proliferation in the evolutionary system. 
As the main novelty of the present work we study the feasibility of $k$-core percolation \cite{Goltsev06} 
as a selection mechanism.  $k$-core percolation is 
a systematic, iterative procedure where a node in a network is removed from a network if it 
sustains less than a fixed number of $k$ links to other nodes. 
We show that evolutionary systems which grow according to a catalytic set dynamics combined with a 
$k$-core selection mechanism, reproduce the observed power law behaviour in good agreement 
with the fossil data
% for all of the 
for the three observables: size of extinction events, lifetime and number of species per genus. 
The model explicitly describes the origination of species and their interactions, 
the fitness landscape is co-evolving with the topology specified by these interactions.

\section{The Model}

In the following species are represented as nodes in a network. We introduce two 
types of links between these nodes, the first type keeping the ancestral relations, 
the other type describing the interactions between species. These links are 
recorded in two separate adjacency matrices, as described below.

\subsection{Growth}
%/Proliferation/Diversification}

The system is initiated with a small number $N_0$ of species.  
These are assumed to be constantly present i.e. they are not subject to the selection mechanism.
New species (nodes) are introduced as mutations of already existing ones. 
They will prove viable only if they receive some "support" from other species. 
At each time step a species may be subject to a mutation which leads to a new node. 
The mutation is favored/suppressed through the influence of other already existing species. 
We identify the probability for the occurrence of a viable mutation  with the effective growth rate $\lambda$ of the system. 
In the absence of any selection mechanisms or extinctions the system diversity  grows 
according to $N\left(t\right)=N_0 \mathrm e^{\lambda t}$.
To take into account ancestral relationships we introduce the  \emph{ancestral  table} $\alpha$, 
a three dimensional tensor with entries $\alpha_{ijk} \in \{0,1\}$. 
Suppose that  species $i$ mutates and gives rise to a new species $k$ and that species $j$ provides 
support for the survival of  $k$. In this case the  ancestral adjacency matrix element $\alpha_{ijk}=1$, otherwise 
$\alpha_{ijk}=0$. Each species is associated  to a genus. If species $i$ is from genus $g_i$, its mutant 
$k$ will most likely be assigned to the same genus $g_i$. However, with a small probability $p^{gen}$ 
the mutation will be large enough that $k$ constitutes a new genus $g_k \neq g_i$. 
The results will, as discussed later, only marginally depend on the actual choice of $p^{gen}$, 
we worked with a figure of $p^{gen}=0.005$.

On top of this ancestral relationship, a new species will also interact with other species in its surrounding. 
The environment of a new species -- its ecological context -- will be strongly determined by the 
environment of its ancestors, i.e. the species the ancestors interact with. 
A given species $k$ (descending from $i$) will thus be most likely to interact with more or less the same 
species as $i$.  $k$ receives a given fraction of interaction-links from $i$. 

As a consequence of this growth rule with the particular copying mechanism, 
clusters of strongly interconnected, interacting species naturally 
emerge. In other words, species in a cluster are highly adapted to each other and form an 
environment to which can be referred to as an "ecological niche".  

Interspecies dependencies are encoded in the interaction matrix $I_{ij}$. 
A general choice for the entries in $I$ would be to introduce a probability that an entry is non-zero, 
i.e. there is an interaction between species $i$ and $j$, and in this case let the values of $I$ vary between 
$-1$ and $+1$, for inhibitive and stimulating influences. 
Evolutionary dynamics of this kind has been studied \cite{Jain01} and it has been shown to 
lead to a proliferation of predominantly stimulating influences (positive entries). 
Thus, since here we are interested in a model for macro-evolution and not ecology, we 
assume only positive binary entries in $I$, i.e. $I_{ij} \in \{0,1\}$ for no interaction, or stimulating influence, respectively.

For later use, the indegree of node $i$, $\kappa_{i}^{in}$ is defined as the sum of 
%all non-zero entries in 
the 
$i$-th line of the interaction matrix $I$, i.e. $\kappa_{i}^{in}=\sum_{\{x \in N\left(t\right)\}}I_{xi}$. Note that 
the number of species $N(t)$, and thus matrices $\alpha$ and $I$ are non-constant over time. 

\begin{figure}[tb]
\begin{center}
\hspace*{-1.0cm}\includegraphics[height=28mm]{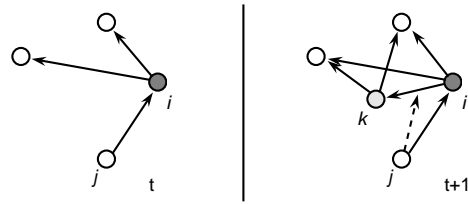}
\end{center}
\caption{At time $t$ (left panel) node $i$ is chosen to mutate. It has one incoming 
link from $j$ and two outgoing links. At time $t+1$ (right panel) node $k$ 
came into existence through the mutation of $i$ under the supportive influence of $j$. 
Here $m=1$, i.e. $k$ copies all outgoing links from $i$.}
\label{copy}
\end{figure}

\subsection{Growth dynamics}

The model consists of a two-step process: a growth and diversification process, followed by a selection procedure. 
During one time step we apply the following procedure to each node in a random update:

\begin{itemize}
\item Pick a node $i$. With probability $1-\lambda$ (same for all $i$) do nothing and pick another node, otherwise 
with probability $\lambda$ do the following:
\item Choose at random one of the nodes linking to $i$, say node $j$. Add a new node $k$ to the network which is a mutation of either $i$ or $j$. Set either $\alpha_{ijk}=1$ or $\alpha_{jik}=1$ with equal probability.
This means that either $i$ or $j$ has mutated.
\item Let us assume $i$ mutated. Then the new species $k$ receives an incoming link from $i$ ($I_{ik}=1$) 
and copies each outgoing link from $i$ with a probability $m$, i.e. if $i$ links to $i'$ ($I_{ii'}=1$), 
$k$ links to $i'$ with probability $m$.  (If it links we set $I_{ki'}=1$).
\item With probability $p^{gen}$ the new species $k$ constitutes a new genus, otherwise $k$ is 
associated with the same ancestor genus $i$. 
\end{itemize}

Effectively, we employ a 'copying mechanism', where a node $i$ gets copied (produces node $k$) 
together with the two types of links involved: 
In the case of the ancestral relationships either a link to $i$ is established or, 
with same probability, one incoming link of $i$, namely from $j$, is copied. 
In the case of the species interactions each outgoing link from $i$ is copied to 
$k$ with probability $m$. See Fig. \ref{copy} for an illustration. A  copying 
mechanism of this kind has been studied in \cite{Vazquez03} 
and was applied
in the context of 
protein interaction networks \cite{Vazetal03}.

\subsection{Selection as k-core pruning}

By assuming that selection predominantly  acts on species of low fitness, 
a quantitative measure for  fitness is necessary. It was argued that a species' individual 
fitness should be  related to the number of stable relationships that this species is able to maintain 
in its environment \cite{deDuve05}. The higher this number, the more interactions ensure its 
survival. In this view one can directly identify the indegree of species 
$i$, $\kappa_i^{in}$,  with its fitness; one can picture $\kappa_i^{in}$ as the total 'support' $i$ gets from its surroundings. 
In this view it is natural to  implement the selection procedure in the following way:  

Suppose there exists an exogenous stress level for all species, $k^{stress}$
which fluctuates due to abiotic causes. 
It can be modelled as a random process drawn at each time step from a Poisson distribution 
$\mathrm{Pr} \left( k^{stress}=n \right)=\left(\theta^n \mathrm{e}^{-\theta} \right)/n!$. 
The mean $\theta$ of this distribution gives the average biotic stress in the system. 
Species with $\kappa_i^{in}<k^{stress}$ become removed from the network with all their links. 
As soon as these nodes are removed some of the surviving nodes will now have an indegree 
smaller than $k^{stress}$ and become extinct too, and so on. 
In other words, at each time step only the $k$-core of the network survives, 
the network is pruned down to its $k$-core.

\section{Theoretical estimates}

We now estimate the distributions of three quantities which are observable in fossil data, 
extinction events, lifetime and species per genus. These are known to be compatible with  
power-law distributions,  with exponents between $1.5$ and $2$ \cite{Newman03}. 
We analytically derive the  exponents for extinction size $\gamma_E$, 
number of species per genus $\gamma_S$, and lifetimes $\gamma_L$, 
and discuss parameter (in)dependence of the results.  
We then compare them to simulations at the end of this section.

\begin{figure}[tb]
\begin{center}
\includegraphics[height=50mm]{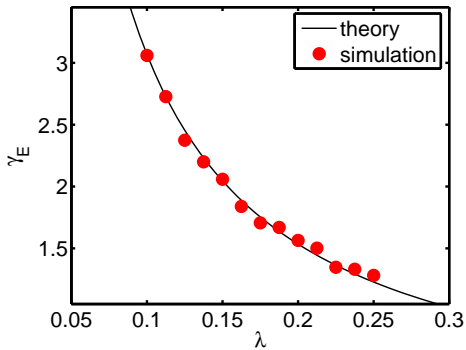}
\end{center}
\caption{We compare the 
prediction from Eq. (\ref{gammaE}) (solid line) with simulation data (red circles) for $\theta=1$.}
\label{ge}
\end{figure}

\subsection{Size distribution of extinctions}

We are interested in the number of species becoming extinct in each time step, 
i.e. the distribution of extinction sizes. It can be derived analytically by making some simplifying 
assumptions. 
A node $i$'s indegree is given by $\kappa_{i}^{in}=\sum_{\{x \in N\left(t\right)\}}I_{xi}$.
Since each node receives an incoming link from its ancestor, the minimal indegree in the network is one.
Thus each species can survive if we prune the network with 
$k^{stress} \in \{0,1\}$. 
The probability $p_{surv}$ for the occurrence of a stress level $k^{stress}$, which does not lead to a single extinction event, is given by a Poisson process  
$p_{surv}=\sum_{n=0}^1 \left(\theta^n \mathrm e^{-\theta}\right)/n! = \mathrm e^{-\theta} \left(1+\theta\right)$. 
With probability $p_{surv}$ the diversity proliferates as $N\left(t+1\right)=N\left(t\right) \left(1+\lambda\right)$. 
We assume that the main contribution to extinction sizes stem from percolation with $k^{stress}=2$, 
which is the case for reasonable choices of the parameters $\lambda$ and $\theta$. 
By reasonable choices we mean values for $\lambda$ and $\theta$ where a nontrivial interplay between the growth and extinction dynamics can \emph{de facto} be observed. Otherwise, keeping $\lambda$ fixed and choosing $\theta$ too low the system would just grow exponentially, conversely for too high $\theta$ all species would vanish within a few iterations. 
We further make the simplifying assumption that if $k^{stress}>1$ occurs, 
a constant fraction $c$ of the entire population will go extinct. 
Thus from our assumptions follows the Ansatz that with probability $1-p_{surv}$ the diversity 
behaves like $N\left(t+1\right)=N\left(t\right) \left(1-c\right)$.

Let us call the number of species becoming extinct at each time step 
$\Delta N^\dagger\left(t\right)$ and assume that $\Delta N^\dagger\left(t\right)=cN\left(t\right)$. 
Then we have $\Delta N^\dagger\left(t\right)/N_0= \exp \left(\lambda t\right)$, 
or equivalently $t=\left(1/\lambda\right) \ln \left(\Delta N^\dagger\left(t\right)/N_0 \right)$. 
Assume that an extinction event occurs at time $t+1$. The probability that the 
system has proliferated over the past $T$ iterations is given by $p_{surv}^T$ ($T$ being an exponent), 
so the probability to find a specific extinction size $\Delta N^*$ is given by 
$\mathrm{Pr} \left(\Delta N^\dagger\left(t\right)=\Delta N^*\right)=p_{surv}^{\left(1/\lambda\right) \ln \left(\Delta N^*\left(t\right)/N_0 \right)}$. 
Taking the natural logarithm on both sides and plugging in for $p_{surv}$ we finally have 
$\ln\left(\mathrm{Pr} \left(\Delta N^\dagger\left(t\right)=\Delta N^*\right) \right) = \mathrm{const}+ \left[\left(-\theta+\ln\left(1+\theta\right) \right)/\lambda\right] \ln\Delta N^*$. 
Thus the distribution of extinction sizes follows a power-law with exponent $\gamma_E$ depending on $\lambda$ and $\theta$:
\begin{eqnarray}
  \mathrm{Pr} \left(\Delta N^\dagger\left(t\right)=\Delta N^*\right)  \propto  \left(\Delta N^*\right)^{-\gamma_E} \quad,
  \\
  \nonumber
  \gamma_E  =  \frac{\theta-\ln\left(1+\theta\right)}{\lambda}
\quad.
\label{gammaE}
\end{eqnarray}
A comparison between this prediction and simulation results from the full model (without assumptions) is shown in Fig. \ref{ge}  for $\theta=1$,
revealing excellent agreement. 
Slopes from the simulation data were estimated using a maximum likelihood 
method \cite{Clauset07}, standard deviations are smaller than symbol size.
This \textit{a posteriori} justifies our simplifying assumption $\Delta N^\dagger\left(t\right)=cN\left(t\right)$.  
The difference to simulation data stems from the fact that also percolations with 
higher $k$ occur albeit exponentially less likely. 
%Although these extinctions contribute negligible to the distribution of extinction sizes, 
%they may introduce mass extinctions, i.e. very rare events where nearly the entire population 
%becomes extinct. Note that the mere existence of the power-law is independent 
%of the actual form of the distribution giving $k$, as becomes apparent 
%from Eq.\ref{gammaE} where only the functional form of $\gamma_E$ would change.  

\subsection{Distribution of species per genus}

Whereas the extinction-size distribution displays explicit parameter on $\lambda$ and $\theta$
dependence, this will be shown to be not the case for the distributions of species per genus and lifetimes. 
Let us start with the indegree distribution of our growth model $p\left( \kappa_i^{in} \right)$ encoded in $I$, 
which is known to be scale-free  \cite{Vazquez03}. Growing networks have  
scale-free degree distribution if they incorporate preferential attachment. 
How is  preferential attachment present in the present model? Consider the avenue of a new species 
$k$ due to a mutation of $i$ under the supportive influence of $j$ and a randomly chosen, already existing node $l$. 
%WER IST $l$???
What is the probability that the indegree of $l$, $\kappa_l^{in}$, will increase by one? 
This can happen if $l$ receives an incoming link from $k$ because it already has an incoming link from 
$i$ which happens with a probability proportional to the indegree of node $l$. 
This introduces preferential attachment and the resulting indegree distribution, 
as worked out in \cite{Vazquez03}, follows a power law with
\begin{equation}
p\left( \kappa^{in} \right) \propto \kappa^{in^{-2}}
\quad,
\label{indeg}
\end{equation}
as long as the link-copying probability  $m>0.4\left(1\right)$ \cite{Vazquez03}
%????
, which we assume to hold. 

Suppose our system size is $N$ species. Denote the number of genera containing $n_s$ species by $\bar n_g(n_s,N)$. It is then straight forward to derive the growth equation
\begin{eqnarray}
\bar n_g(n_s, N+1) = \bar n_g(n_s,N) -\bar n_g(n_s,N) w(n_s)+ \\
\nonumber
+ \bar n_g(n_s-1,N) w(n_s-1)
\quad,
\label{growtheq}
\end{eqnarray}
where $w(n_s)$ is the probability for each genus of size $n_s$ to increase its size by one. The dependence on $p^{gen}$ is introduced in the boundary conditions given by $\bar n_g(n_s=1,N+1)$. For each node associated to an already existing genus, the number of $p^{gen}/(1-p^{gen})$ nodes are added to this one per time step, so we get $\bar n_g(1,N+1)=\bar n_g (1,N)-w(1)\bar n_g(1,N)+p^{gen}/(1-p^{gen})$.  We are interested in stationary solutions of Eq. \ref{growtheq}, i.e. solutions which are independent of the actual system size $N$. For this let us define $n_g(n_s)\equiv N \bar n_g(n_s,N)$. The probability for a genus of size $n_s$ to increase its size by one is obviously $w(n_s)=n_s/N$, this can be interpreted as the probability that a new node copies the genus information from a node of a genus of this respective size. Plugging all this into Eq. \ref{growtheq} we get the recursive relationship $n_g(n_s)=[(n_s-1)/(n_s+1)] \cdot n_g(n_s-1)$ from which one can readily conclude  $n_g(n_s)=f(p^{gen}) \cdot (n_s(n_s+1))^{-1}$ where $f(p^{gen})$ is a constant, thus we have $n_g(n_s) \propto n_s^{-2}$ to leading order.

An important feature of $k$-core percolation is that it preserves statistical invariants 
\cite{CorMur07}, that is, if the original network follows a scale-free degree distribution with a 
given exponent, its $k$-core has the same distribution up to the cut-off at $k$. 
The scale-free network architecture imposed by our growth and diversification rules 
will not be altered by extinction events. Thus the species per 
genus distribution of the model is
\begin{equation}
  n_g\left(n_s\right) \propto n_s^{-2}
\quad, 
\label{spg}
\end{equation}
i.e. $\gamma_S=2$, for the distribution of taxon sizes.

\begin{figure*}[tb]
\begin{center}
\includegraphics[height=44mm]{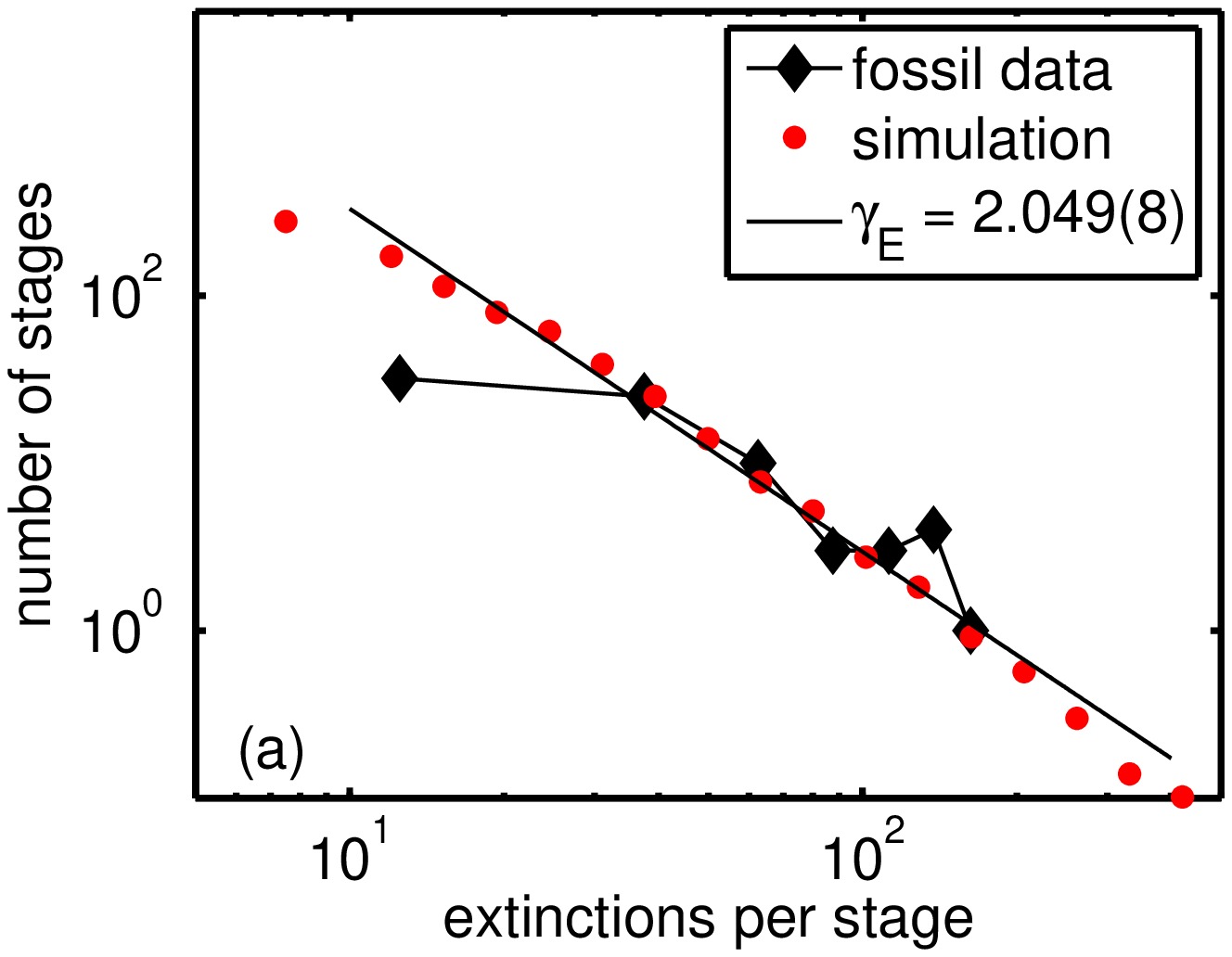}
\includegraphics[height=44mm]{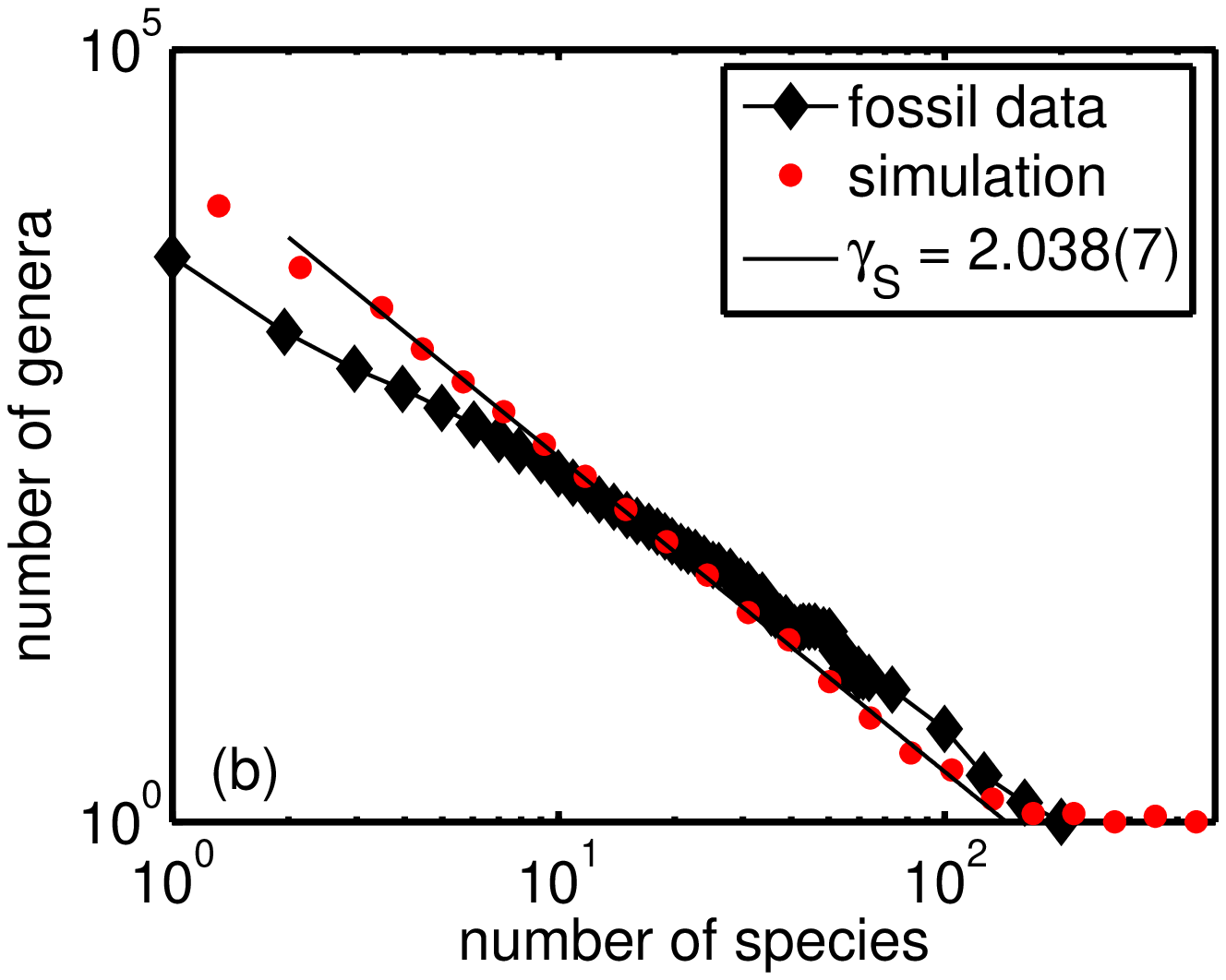}
\includegraphics[height=44mm]{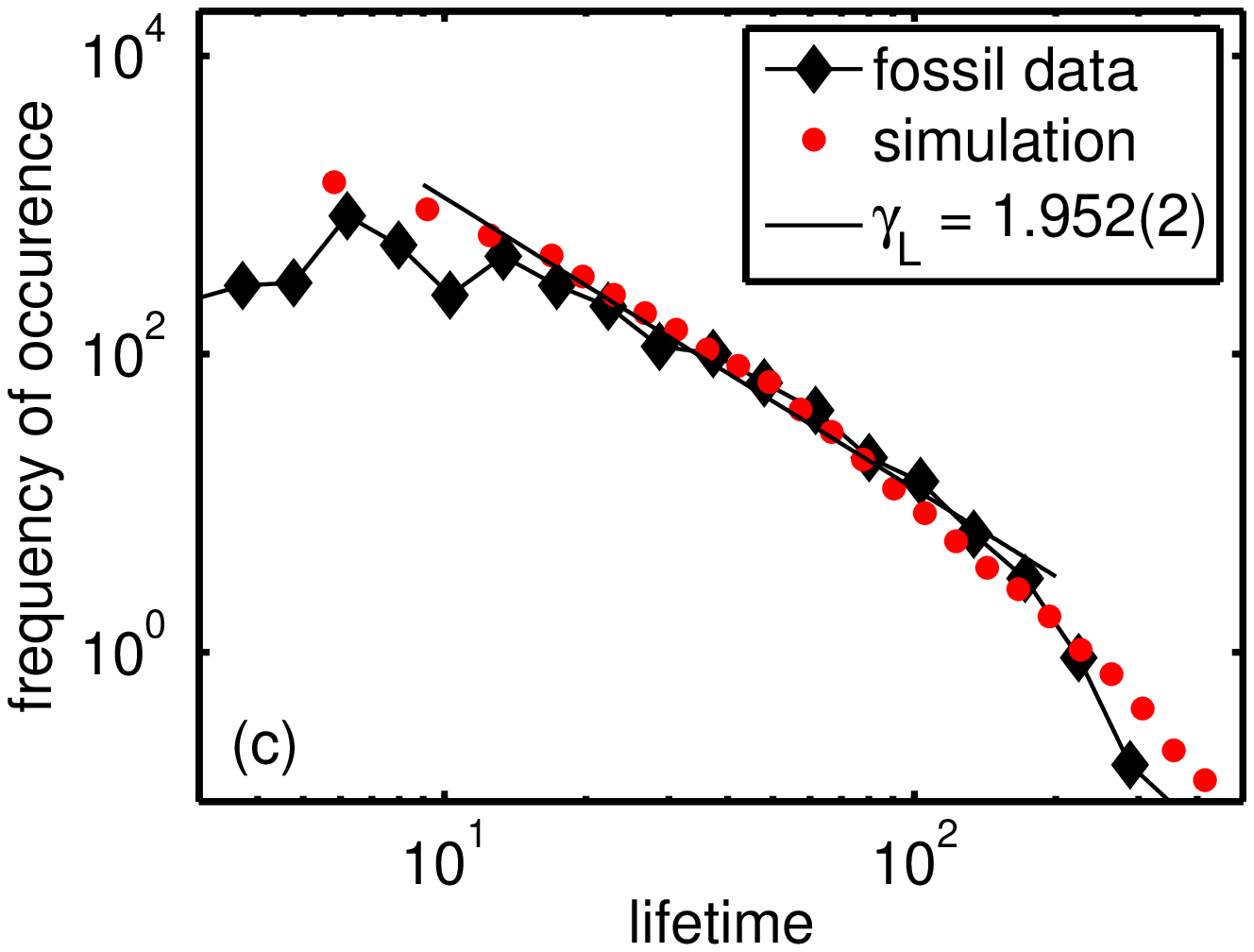}
\end{center}
\caption{Comparison of 
fossil data \cite{Willis22, Sepkoski92} (solid line with diamonds) with simulation data 
(red circles) for the three observables (a) extinction event size, (b) species per genus and (c) 
lifetimes. The straight line is the maximum likelihood estimate for the power-law exponent 
of the simulation data and indicates the range where the fit was applied.}
\label{datplots}
\end{figure*}

\subsection{Lifetime distribution}

To  estimate for the distribution of lifetimes of long living species, i.e. species which will 
not become extinct after the first few iterations, we ask for the lifetime $\tau_i$ of species 
$i$ with an indegree $\kappa_i^{in}$ drawn from $p\left(\kappa_i^{in}\right)$. 
The probability that the stress level will be higher than the node's indegree is 
$\mathrm{Pr} \left( k^{stress}>\kappa_i^{in}\right)=\sum_{k=\kappa_i^{in}+1}^{\infty} \left(\mathrm e^{-\theta} \theta^k\right)/k!$. 
The leading term in this sum is $k=\kappa_i^{in}+1$. 
Consider that  $T$ iterations of the dynamics have taken place. 
We are asking for long lived species, i.e. that within $T\gg1$ iterations there occurs no stress level 
higher than $\kappa_i^{in}$.  Generally, the probability that within $T$ trials with success probability $\mathrm{Pr} \left( k^{stress}>\kappa_i^{in}\right)$ zero successes are obtained is given by a binomial distribution. For large sample sizes $T$ the binomial distribution approaches a Poisson distribution, independent of $T$. Accordingly, in our case the probability for zero successes (the occurrence of no stress level $k^{stress}>\kappa_i^{in}$) follows a Poisson distribution  $\mathrm e^{-\mathrm{Pr} \left( k^{stress}>\kappa_i^{in}\right)}$.
From this it can be concluded that the probability to encounter a species $i$ with lifetime 
$\tau$, i.e.,  $\mathrm{Pr}\left(\tau_i=\tau\right)$, can be estimated from the node's indegree only. 
%It is then given by the probability that a node has a certain indegree $\kappa_i^{in}$??? and simultaneously that  the  
%stress level is never higher than $\kappa_i^{in}$???  for $T\gg1$ iterations, 
One can identify a necessary criterion for the survival of a node, namely that it has an indegree $\kappa_i^{in}$ which is not exceeded by the stress level $k^{stress}$ for $T\gg1$ iterations. So the probability to encounter a lifetime $\tau$ is given by the probability for the occurrence of a stress level higher than $\kappa_i^{in}$,
\begin{equation}
  \mathrm{Pr}\left(\tau_i=\tau\right) \propto p\left(\kappa_i^{in}\right) \mathrm e^{\mathrm{Pr} \left( k^{stress}>\kappa_i^{in}\right)}
  \quad.
\label{lt1}
\end{equation}
The probability to find a node with lifetime $\tau$ is proportional to the probability of finding 
a node with a given indegree $\kappa^{in}$, truncated with the probability for the occurrence of 
specific stress levels. There exists a regime where 
$\mathrm{Pr}\left(\tau_i=\tau\right) \propto \mathrm{Pr}\left(\kappa_i^{in}=\kappa^{in}\right)$ 
holds and by virtue of Eq. (\ref{indeg}) we find $\gamma_L=2$, i.e.
\begin{equation}
   \mathrm{Pr}\left(\tau_i=\tau\right) \propto \tau^{-2}    \quad.
\label{lt2}
\end{equation}

\subsection{Simulations}

We compare simulation results of the presented model to fossil data for extinctions and lifetime drawn from \cite{Sepkoski92},
as well as species per genus after \cite{Willis22} in Fig. \ref{datplots}. 
The model was implemented in a MatLab program and executed until a statistics of $2\cdot 10^4$ extinction events 
were  accumulated. This corresponds to sample sizes of $10^5-10^6$ for individual lifetimes and 
numbers of species per genus, depending on the parameter settings. 
For $\lambda p_{surv}<c\left(1-p_{surv}\right)$ the size of the network does not 
diverge over time and the samples can be obtained from a single run of the simulation. 
For $\lambda p_{surv}>c\left(1-p_{surv}\right)$ the system tends to grow infinitely large; 
for practical purposes we aborted runs as soon as $N\left(t\right)>10^4$ and 
iterated until a satisfactory statistics was reached.

For the extinction events the resulting slope is parameter dependent, we used the setting 
$\left(\lambda=0.15,m=1,\theta=1\right)$ to obtain agreement with the slope of $\gamma_E=2.0\left(2\right)$ 
from the fossil data \cite{Raup91}. For the number of species per genus and lifetimes the 
distributions are independent of the parameter settings and given by the topology (which is a scale-free indegree distribution for values of $m>0.4(1)$) of our 
network only, yielding $\gamma_S=2, \gamma_L=2$. For all three subplots the simulation data was
fitted with a maximum likelihood estimation \cite{Clauset07}, the range of the fit is indicated 
by the range of the straight line. Subsequently the numerical results were binned 
logarithmically and, if necessary, shifted multiplicatively to enhance the clarity 
of the plots. 
Our exponents are summarized and compared to several previous models in Table \ref{tab:expo}.  

\begin{table}[b] 
\caption{Exponents of the distributions of extinction sizes $\gamma_E$, species per genus $\gamma_S$, and 
lifetimes $\gamma_L$, as obtained from the fossil record and 
compared to the exponents of various well known evolution models. The value for $\gamma_E$ from this model was obtained from simulations with $\lambda=0.15, m=1, \theta=1$.}
\label{tab:expo}
\vspace{10pt}
\centerline{
\begin{tabular}{|l|ccc|}
\hline
& $\gamma_E$ & $\gamma_S$ & $\gamma_L$ \\
\hline
fossil data &2.0(2) &1.7(3) &1.5(1) \\
\cline{2-4}
Kauffman &$\simeq$1 &- &- \\
Bak and Sneppen &1 to 3/2 &1 &- \\
Sol\'e and Manrubia &2.05(6) &- &2.05(6) \\
Newman & 2.02(2)& 1.03(5)&1.6(1) \\
present model & 2.049(8) &2 &2 \\ 
\hline
 \end{tabular}}
\end{table}

\section{Discussion}

We presented a 
model for evolution which reproduces statistical features observed in 
fossil data. An evolutionary system is modelled as a catalytic network 
with two superimposed  network topologies, one  incorporating species-species interactions, 
the other the phylogenetic tree structure. The fitness of species is given by the 
connectivity structure of the network, thus naturally a co-evolving fitness landscape arises.
Fitness becomes nothing but a co-evolving topological entity, the more relationships a species is able to build and sustain, the fitter it becomes. 
Species interactions are introduced by a variant of preferential attachment known as 
'copying mechanism' \cite{Vazquez03}.  
Without any further assumptions this mechanism leads to a natural emergence of "ecological niches", 
which in network terms relate to a high degree of  clustering in the network.

In this model we have taken a gradualist viewpoint concerning speciation in assuming 
that the growth rate $\lambda$ is constant. However, this choice was only made for reasons of 
simplicity. Benton and Pearson \cite{Benton01} propose that gradual speciations are more likely to 
occur in stable environments (as it is the case for e.g. marine plankton), whereas marine 
invertebrates and vertebrates are more likely to show a punctuated pattern of speciation. 
The latter case could be naturally  introduced in our model by assuming a functional dependence 
$\lambda \equiv \lambda \left( \theta \right)$, i.e. introducing a mechanism that couples the growth rate with 
the actual values of $k^{stress}$. Irrespective of this choice, the main characteristics 
of our model would not be altered. The number of species per genus and lifetimes only 
depends on topological features of the network which would not be affected by a varying 
growth rate. Our analysis for the extinction sizes would hold too, except that one has  
to set $\lambda= \lambda \left( \theta \right)$ in Eq. (\ref{ge}). The existence of the power law 
is independent of both the functional form of the growth rate and the stochastic stress level. 

Our selection mechanism differs from the one studied by Sol\'e and Manrubia \cite{Sole96} in that extinction avalanches spread over successive time steps in their model and that each species becoming extinct is immediately replaced by a randomly chosen one (therefore leading rather to a model for ecology where empty niches are re-filled), whereas in our model the selection mechanism acts on a 'snapshot' of the population and does not depend on which randomly chosen species replaces an extinct one. Our pruning procedure further differs from the selection mechanism adopted by Newman \cite{Newman03} in that each species has a randomly assigned fitness value (independent of interspecies relationships) and species below a given stress level become extinct, which is contrasted by mass extinctions of causally connected species in our model.   

We suggested the use of $k$-core percolation as a mechanism to select species according 
to their fitness values. On a technical level this allows to understand the system by studying 
its $k$-core architecture. If the applicability of this mechanism to prune the 'tree of life' can be justified beyond the statistical features presented here, remains an open question.

\end{document}